\documentclass[12pt]{article}
\usepackage[russian]{babel}
\usepackage{amssymb}
\usepackage{multirow}
\usepackage{enumerate}
\usepackage{hhline}
\usepackage{amsmath}
\usepackage{amsthm}
\usepackage{listings}
\usepackage{graphicx}
\usepackage[table,usenames,dvipsnames]{xcolor}
\usepackage{url}
\usepackage[unicode,bookmarksopen,bookmarksopenlevel=1,bookmarksnumbered]{hyperref}
\hypersetup{pdfstartview={XYZ null null 1.25}}

\newtheorem{theorem}{Theorem}

\def\hcorrection{\hspace{-0.3em}}

\def\Author#1{\vspace{4.0ex plus 0.2ex minus 0.2ex}\centerline{\Large{#1}}}
\def\Title#1{\section*{\hcorrection{#1}}}

\def\References#1{{\footnotesize\baselineskip=12pt\begin{thebibliography}{99}{#1}\end{thebibliography}}}

\def\myf#1{\mathit{\tilde{#1}}}
\def\myff#1{\mathit{\tilde{\tilde {#1}}}}
\def\sps{, \ \ }
\def\spsd{. \ \ }
\begin{document}
\renewcommand{\proofname}{Proof.}
\renewcommand{\refname}{References}
\renewcommand{\figurename}{Fig.}
\renewcommand{\contentsname}{Contents}
\renewcommand{\tablename}{Table}
\renewcommand{\listtablename}{Tables}
\renewcommand{\listfigurename}{Figures}
\lstset{language=Delphi,basicstyle=\tiny,commentstyle=\color{green}}
\lstset{numbers=left, numberstyle=\tiny, stepnumber=1, numbersep=5pt,formfeed=\newpage}
%%%%%%%%%%%%%%%%%%%%%%%%%%%%%%%%%%%%%%%%%%%%%%%%%%%%%%%%%%%%%%%%%%%%%%%%%%%%%%%%
\foreignlanguage{english}{
\begin{center}
\Title{Sketches from the life of hypercomplex numbers.}
\Author{K.V. Andreev}
\end{center}

Every science result represents an element of a puzzle from the general picture of the world. Every element has connectors with the help of which it connected to its neighbors. A sketch is called a description of how these connectors are connected to each other. But each painter represents the world as he sees it. Therefore, all pictures of the world are different. And all elements of a puzzle are connected to each other whimsically. This article is a description of the small piece of a full picture from the hypercomplex life as it is seen to the author. Therefore, in the article, the main ideas of the induction construction \href{http://arxiv.org/abs/1204.0194}{arXiv:1204.0194}, \href{http://arxiv.org/abs/1110.4737}{arXiv:1110.4737}, \href{http://arxiv.org/abs/1202.0941}{arXiv:1202.0941}, \href{http://arxiv.org/abs/1208.4466}{arXiv:1208.4466} are considered in the form of sketches. By and large, the article establishes a link between Clifford algebras and alternative-elastic algebras at the level of connectors.\\

\begin{enumerate}
\item \textbf{Real numbers.}

\hspace{0.5cm} The first axiom system for arithmetic of real numbers was published by Hilbert in 1900. Real numbers can be determined by the following axioms:
\begin{enumerate}[I.]
\item Field Axioms.
\begin{enumerate}[{I}-1.]
\item Closure of $\mathbb R$ under addition: $\forall a,b\in \mathbb R,\ \exists !\ a+b=:c \in \mathbb R$.
\item Associativity of addition: $\forall a,b,c\in \mathbb R,\ a+(b+c)=(a+b)+c$.
\item Commutativity of addition: $\forall a,b\in \mathbb R,\ a+b=b+a$.
\item Existence of additive identity element: $\forall a\in \mathbb R,  \exists\ 0 \in \mathbb R, a+0=a$.
\item Existence of additive inverse: $\forall a\in \mathbb R,  \exists\ b \in \mathbb R, a+b=0$.
\item Closure of $\mathbb R$ under multiplication: $\forall a,b\in \mathbb R,\ \exists !\ a\cdot b=:c \in \mathbb R$.
\item Associativity of multiplication: $\forall a,b,c\in \mathbb R\ a\cdot (b\cdot c)=(a\cdot b)\cdot c$.
\item Commutativity of multiplication: $\forall a,b\in \mathbb R,\ a\cdot b=b\cdot a$.
\item Existence of multiplicative identity element: $\forall a\in \mathbb R,  \exists\ 1 \in \mathbb R, a\cdot~1~=~a$.
\item Existence of multiplicative inverse: $\forall a\in \mathbb R,\   \exists\ b \in \mathbb R,\ a\cdot b=1$.
\item Distributivity of multiplication over addition: $\forall a,b,c\in \mathbb R\ a\cdot (b+c)=$ $=(a\cdot b)+(a\cdot c)$.
\item Nontriviality: $1\ne 0$.
\end{enumerate}
\item Ordering Axioms.
\begin{enumerate}[{II}-1.]
\item Reflexivity: $\forall a\in \mathbb R,\ a\le a$.
\item Antisymmetry:$\forall a,b\in \mathbb R,\ (a\le b)\wedge(b\le a)\Rightarrow(a=b)$.
\item Transitivity:$\forall a,b,c\in \mathbb R,\ (a\le b)\wedge(b\le c)\Rightarrow(a\le c)$.
\item Totality:$\forall a,b\in \mathbb R,\ (a\le b)\vee(b\le a)$.
\item Relationship between order and addition: $\forall a,b,c\in \mathbb R,\ (a\le b)\Rightarrow$ $\Rightarrow(a+c\le b+c)$.
\item Relationship between order and multiplication: $\forall a,b,c\in \mathbb R$,\\ $(0\le a)\wedge(0\le b)\Rightarrow(0\le ab)$.
\end{enumerate}
\item Completeness Axiom.\\
If $A$ and $B$ are nonempty subsets of $\mathbb R$ such that for every $a\in A$ and for every $b\in B$ we have $a\le b$, then there exists $c\in R$ such that $a\le c\le b$ for all $a\in A$ and all $b\in B$.
\end{enumerate}
Note that in 1936, Alfred Tarski set out an axiomatization of the real numbers, consisting of only the 9 axioms \cite[p. 275 (rus)]{Tarski1}.
\item \textbf{Complex numbers.}

\hspace{0.5cm}Complex numbers were introduced by Italian mathematician Gerolamo Cardano in the 16th century (1545). A complex number is represented as $a+bi$, where i is called the imaginary unit and $i^2 = -1$. $a$ and $b$ are real numbers. Unlike real numbers, complex numbers don't form an ordered field. The complex number $a-bi$ is called \textit{complex conjugate} to $a+bi$.

\item \textbf{Cayley-Dickson numbers.}

\hspace{0.5cm} Let an algebra $\mathbb A$ be given then the algebra $\mathbb A^2$ is determined as follows from \cite{Albert1} (in 1942). For all $a,b\in \mathbb A$ and the new imaginary unit $i$, the hypercomplex numbers $x:=a+bi$ form $\mathbb A^2$ and the parities
\begin{enumerate}[1.]
\item $a(bi)=(ba)i$;
\item $a(ib)=(a\bar b)i$;
\item $(ai)(bi)=-\bar ba$;
\item $\overline{a+bi}=\bar a-bi$
\end{enumerate}
are executed.

\item \textbf{Quaternions.}

\hspace{0.5cm} Quaternions were introduced by Hamilton in 1843 \cite{Hamilton1} with the three imaginary units ($ij=k,\ jk=i,\ ki=j$). However, the quaternions form an non-commutative associative division algebra. Each quaternion is formed by a pair of complex numbers according to the Cayley-Dickson procedure.

\item \textbf{Octonions.}

\hspace{0.5cm} Octonions were introduced by John T. Graves (in 1843) and Arthur Cayley (in 1845) independently \cite{Cayley1}, \cite{Baez1} with seven imaginary units. However, the octonions form an non-commutative non-associative alternative division algebra. Each octonion is formed by a pair of quaternions according to the Cayley-Dickson procedure. Instead of the associative low, the alternative lows exists: $(aa)b=a(ab)$ (left), $b(aa)=(ba)a$ (right). The automorphism group has the dimension equal to 14 and is a subgroup of the orthogonal group whose the dimension is equal to 28. One of the invariants is the algebra identity that reduces the dimension on 7. What is the second invariant that reduces the dimension on 7 else?
\newpage

\item \textbf{Sedenions.}

\hspace{0.5cm} Sedenions were introduced by John T. Graves with 15 imaginary units. However, the sedenions form an non-commutative non-associative alternative-elastic non-division algebra. Each sedenion is formed by a pair of octonions according to the Cayley-Dickson procedure. Instead of the alternative low, the weakly alternative low exists: $(aa)b-a(ab)=b(aa)-(ba)a$. This identity is equivalent to both the flexible low $a(ba)=(ab)a$ and the power-associative low a(aa)=(aa)a \cite{Albert1}, \cite{Moreno1}, \cite{Andreev4}.

\item \textbf{Alternative-elastic algebras.}

\hspace{0.5cm} An alternative-elastic algebra $\mathbb A^n$ over field $\mathbb R$ is formed by the following axioms \cite{Andreev4}:
\begin{enumerate}[1.]
\item Closure of $\mathbb A^n$ under addition: $\forall a,b\in \mathbb A^n,\ \exists !\ a+b=:c \in \mathbb A^n$.
\item Associativity of addition: $\forall a,b,c\in \mathbb A^n,\ a+(b+c)=(a+b)+c$.
\item Commutativity of addition: $\forall a,b\in \mathbb A^n,\ a+b=b+a$.
\item Existence of additive identity element: $\forall a\in \mathbb A^n,  \exists\ 0 \in \mathbb A^n, a+0=a$.
\item Existence of additive inverse: $\forall a\in \mathbb A^n,  \exists\ b \in \mathbb A^n, a+b=0$.
\item Closure of $\mathbb A^n$ under multiplication: $\forall a,b\in \mathbb A^n,\ \exists !\ a\cdot b=:c \in \mathbb A^n$.
\item Weakly alternativity of multiplication: $\forall a,b\in \mathbb A^n\ (a\cdot a)\cdot b-a\cdot (a\cdot b)=$ $=b\cdot (a\cdot a)-(b\cdot a)\cdot a$.
\item Existence of multiplicative identity element: $\forall a\in \mathbb A^n,  \exists\ 1 \in \mathbb A^n, a\cdot~1~=~a$.
\item Existence of multiplicative inverse: $\forall a\in \mathbb A^n,\   \exists\ b \in \mathbb A^n,\ a\cdot b=1$.
\item Distributivity of multiplication over addition: $\forall a,b,c\in \mathbb A^n\ a\cdot (b+c)=(a\cdot b)+(a\cdot c)$.
\item Nontriviality: $1\ne 0$.
\end{enumerate}

One can determine the metric on the algebra as $<a,b>e:=\frac{1}{2}(a\bar b+b\bar a)$. In particular, each Cayley-Dickson algebra, constructed from $\mathbb R$ by Cayley-Dickson procedure applied a number of times,  is a metric alternative-elastic algebra.

\item \textbf{Complex and real representations. Real inclusion.}

\hspace{0.5cm}  In the article \cite{Norden1} (in 1959), the Norden affinor was introduced as $\Delta_\myf\Lambda{}^\myf\Psi~:=$\\$=\frac{1}{2}(\delta_\myf\Lambda{}^\myf\Psi+if_\myf\Lambda{}^\myf\Psi)\ (\myf\Lambda,\myf\Psi,...=\overline{1,2n})$, where $f_\myf\Lambda{}^\myf\Psi$ is a complex structure. In 1989, this allowed to introduce the Neifeld operators \cite{Newfield1} $\Delta_\myf\Lambda{}^\myf\Psi:=m^\Lambda{}_\myf\Lambda m_\Lambda{}^\myf\Psi$ ($\Lambda,\Psi,...=\overline{1,n}$). This provided a transition between the real and complex representations: $\mathbb R^{2n}_{(n,n)}$ and $\mathbb C^n$.
In addition, one can locally define a real inclusion with the help of an inclusion operator $H_i{}^\Lambda$: $\mathbb R^n\rightarrow \mathbb C^n$ according to \cite{Newfield1} and can determine the involution in the complex space as $S_\Lambda {}^{\Psi '}=$ $=H^i{}_\Lambda\bar  H_i{}^{\Psi '}\ (\bar H_i{}^{\Psi '}:=\overline{H_i{}^\Psi}$ - complex conjugation) ($i,j,...=\overline{1,n}$). In particular, for the metric tensors, the connecting parities
$g_{\Lambda\Psi}:=G_{\myf\Lambda\myf\Psi}m_\Lambda{}^\myf\Lambda m_\Psi{}^\myf\Psi\sps  \bar g_{\Lambda '\Psi '}:=$ $=G_{\myf\Lambda\myf\Psi} \bar m_{\Lambda '}{}^\myf\Lambda \bar m_{\Psi '}{}^\myf\Psi\sps$ $g_{ij}:=H_i{}^\Lambda H_j{}^\Psi g_{\Lambda\Psi}=\bar H_i{}^{\Lambda '}\bar H_j{}^{\Psi '}\bar g_{\Lambda'\Psi'}$ are executed \cite[p. 12, eq. (3.11); p. 17, eq. (4.10)]{Andreev0}, \cite[p. 11, eq. (3.11); p. 15, eq. (4.10)(eng), p. 135(12), eq. (3.11); p. 140(17), eq. (4.10)(rus)]{Andreev1}.

\item \textbf{Clifford algebras.}

\hspace{0.5cm} Clifford algebra (was discovered in 1878 by Clifford William Kingdom) $CL(G^{2n}_{(n,n)})$ over the field $\mathbb R$ is determined by generators $\gamma_\myf\Lambda$ ($\myf\Lambda,\myf\Psi,...=\overline{1,2n}$) satisfied the Clifford equation $\gamma_\myf\Lambda\gamma_\myf\Psi+\gamma_\myf\Psi\gamma_\myf\Lambda=G_{\myf\Lambda\myf\Psi}E$, where $E$ is identity operator. Each generator is represented by a matrix whose the dimension is equal to $(2N)^2\times (2N)^2$, where $N:=2^{n/2-1}$. The space $\mathbb R^{2n}_{(n,n)}(\mathbb C^{n})$, on which the generators are defined, is called \emph{base space}. One can pass to one of the $2N$ complex representations such the generators (see next Section) that leads to the complex algebra $CL(g^n)$, each generator of which is represented by a matrix whose the dimension is equal to $2N\times 2N$. A generator of the Clifford algebra $CL(g^{n}_{(n,0)})$ can be constructed with the help of an inclusion operator for the inclusion of the real space $\mathbb R^n$ into the complex space $\mathbb C^n$. Note that the Clifford generators can be reduced by the formula $\gamma_\Lambda=\left(
\begin{array}{cc}
0              & \sigma_\Lambda \\
\eta_\Lambda  & 0
\end{array}\right)$. And the components can be formalized as $\eta_\Lambda:=\eta_\Lambda{}^{AB}\sps \sigma_\Lambda:=\sigma_\Lambda{}_{AB}=\eta_\Lambda{}_{BA};$ $\ (A,B,... =\overline{1,N},\ \Lambda,\Psi,... =\overline{1,n})$. And the Clifford equation will take the form $\eta_\Lambda{}^{AB}\eta_\Psi{}_{CB}+\eta_\Psi{}^{AB}\eta_\Lambda{}_{CB}=g_{\Lambda\Psi}\delta_C{}^A$. Accordingly, the space $\mathbb C^N$ ($\mathbb C^{2N^2}$) is called \emph{spinor space}. The operators $\eta_\Lambda{}^{AB}$ are the connecting ones and define \emph{spinor formalism}. This means that a path of the algebraic load is transferred onto the connecting operators; this simplifies some tensor parities and leads to new spinor ones, which are not obvious in the tensor form.  A good example is the spinor classification of the Weil tensor \cite[v. 2, p. 256(eng)]{Penrose1} unlike the tensor ones \cite{Petrov1}. For our purposes, such an algebraic expression, which carries most of the algebraic load, will be the triple product $\eta_\Lambda{}^{AB}\eta_\Psi{}_{AC}\eta_\Phi{}_{DB}$ \cite[p. 272]{Postnicov1} written down in one form or another. Note that both the connecting operators $\eta_\Lambda{}^{AB}$ and the structural constants of an algebra $\eta_{\Lambda\Psi}{}^{\Phi}$ have three indexes for each. Can one associate these two objects together? And what will be the third object of this link? Will this be the second invariant for the octonion automorphism group that reduces the dimension on 7 else?  The answer to these questions is here \cite{Andreev3}, \cite[p. 48, eq. (9.3)(eng), p. 175(55), eq. (9.3)(rus)]{Andreev1}, \cite[p. 55, eq. (9.3)]{Andreev0}, \cite[eq. (43)-(45)]{Andreev4}, \cite[eq. (2)]{Andreev5}.
$$
\eta_{\Lambda\Psi}{}^\Phi:=\sqrt{2}\eta_\Lambda{}^{AB}\eta_{\Psi CA}\eta^\Phi{}_{DB}\theta^{CD}\sps
$$
where $\theta^{CD}$ is the controlling symmetric spin-tensor and is the second invariant for the octonion automorphism group. For the real inclusion $H_i{}^\Lambda:\mathbb R^{n} \rightarrow \mathbb C^{n}\ (i,j,... =\overline{1,n})$, such an algebra with the structural constant $\eta_{ij}{}^k$ will be a metric alternative-elastic algebra over the field $\mathbb R$. Well, but how to solve the Clifford equation for even n? And why controlling symmetric spin-tensor is an invariant under the octonion automorphism group? We need to look at the history again. But at the beginning, it is necessary to answer on the question: What is a complex representation of the real connecting operators?

\item \textbf{Complex representation of the connecting operators.}

\hspace{0.5cm} The Neifeld operators induce spinor analogues on the spinor space. In this case, every I-th subspace $\mathbb R^4_{(2,2)}\subset R^{2n-4(I-1)}_{(n-2(I-1),n-2(I-1))}\oplus \mathbb C^{2(I-1)}$ has its own Neifeld operator $m_I$ ($N=2^{n/2-1}$, $I=\overline{1,\frac{n}{2}}$) \cite[Algorithm 6.1, p. 77(eng), pp. 154-155(31-32)(rus)]{Andreev1}, \cite[pp. 31-32]{Andreev0}. The Norden affinors $\Delta_I$ induce the two pairs of operators $(\tilde \Delta_I)_\pm,(\myff \Delta_I)_\pm$ on the spinor space  ($\Lambda,...=\overline{1,2n-2I}$, $\myf\Lambda,...=\overline{1,2n-2(I-1)}$, $A,B,C,D ... =\overline{1,\frac{2N^2}{2^I}}$, $\myff A,\myff B,\myff  C,\myff  D ... =\overline{1,\frac{2N^2}{2^{I-1}}}$)
$$
\begin{array}{c}
(\Delta_I)_\myf\Lambda{}^\myf\Psi\eta_\myf\Psi{}^{\myff A\myff B}=\eta_\myf\Lambda{}^{\myff C\myff D}(\tilde\Delta_I)_-{}_\myff C{}^\myff A(\tilde{\tilde\Delta}_I)_-{}_\myff D{}^\myff B+\eta_\myf\Lambda{}^{\myff C\myff D}(\tilde\Delta_I)_+{}_\myff C{}^\myff A(\tilde{\tilde\Delta}_I)_+{}_\myff D{}^\myff B\sps\\
(m_I)_\Lambda{}^\myf\Psi(m_I)^\Lambda{}_\myf\Lambda=(\Delta_I)_\myf\Lambda{}^\myf\Psi\sps\\
(\tilde m_I)_\pm{}_\Lambda{}^\myf\Psi(\tilde m_I)_\pm{}^\Lambda{}_\myf\Lambda=(\tilde \Delta_I)_\pm{}_\myf\Lambda{}^\myf\Psi\sps
(\myff m_I)_\pm{}_\Lambda{}^\myf\Psi(\myff m_I)_\pm{}^\Lambda{}_\myf\Lambda=(\myff \Delta_I)_\pm{}_\myf\Lambda{}^\myf\Psi\spsd
\end{array}
$$
Note that $\Delta_I$ is both the identity operator on $\mathbb R^{2n-4I}_{(n-2I,n-2I)}\oplus \mathbb C^{2(I-1)}$ and the canonical Norden affinor on $\mathbb R^{4}_{(2,2)}$. Let $z_I$ be equal to 0 for $\ll-\gg$ and 1 for $\ll+\gg$ then one can define the pair of operators $\tilde M_K\sps \tilde{\tilde M}_K$
($A,B,C,D ... =$ $=\overline{1,N}$, $\myff A,\myff B,\myff  C,\myff  D ... =\overline{1,2N^2}$, $K,J=\overline{1,2N}$, $K\ne J$)
$$
\begin{array}{cc}
\tilde M_K:=\tilde m_{\frac{n}{2}_{z_{\frac{n}{2}}}}\tilde m_{{\frac{n}{2}-1}_{z_{\frac{n}{2}-1}}}\cdot ...\cdot\tilde m_{2_{z_2}}\tilde m_{1_{z_1}}\sps &
\tilde{\tilde M}_K:=\tilde{\tilde m}_{\frac{n}{2}_{z_{\frac{n}{2}}}}\tilde{\tilde m}_{{\frac{n}{2}-1}_{z_{\frac{n}{2}-1}}}\cdot ...\cdot\tilde{\tilde m}_{2_{z_2}}\tilde{\tilde m}_{1_{z_1}}\sps\\
(\tilde M_K)_C{}^\myff A(\tilde M^*_K)^B{}_\myff A=\delta_C{}^B\sps &
(\tilde{\tilde M}_K)_C{}^\myff A(\tilde{\tilde M}^*_K)^B{}_\myff A=\delta_C{}^B\sps\\
(\tilde M_K)_C{}^\myff A(\tilde M^*_J)^B{}_\myff A=0\sps &
(\tilde{\tilde M}_K)_C{}^\myff A(\tilde{\tilde M}^*_J)^B{}_\myff A=0\sps\\
\sum\limits_{K=1}^{2N}(\tilde M_K)_A{}^\myff C(\tilde M^*_K)^A{}_\myff B=\delta_\myff B{}^\myff C\sps &
\sum\limits_{K=1}^{2N}(\tilde{\tilde M}_K)_A{}^\myff C(\tilde{\tilde M}^*_K)^A{}_\myff B=\delta_\myff B{}^\myff C\\
\end{array}
$$
for the spinor space. Finally, one can obtain the $2N$ complex representations of the connecting operators  $(\eta_K)_\Lambda{}^{AB}:=(\tilde{\tilde M}^*_K)^A{}_{\myff A} (m_\Lambda{}^\myf\Lambda\eta_\myf\Lambda{}^{\myff A\myff B})(\tilde M^*_K)^B{}_{\myff B}$ ($\Lambda,...=\overline{1,n}$, $\myf\Lambda,...=\overline{1,2n}$), but not all of them can be significant. Note that $I_\myf\Lambda{}^\myf\Omega(f_I)_\myf\Omega{}^\myf\Psi$ ($\myf\Lambda,...=\overline{1,2n-2(I-1)}$) is the complex pseudo-orthogonal transformation, where $(f_I)_\myf\Lambda{}^\myf\Psi$ is both the complex structure on  $\mathbb R^4_{(2,2)}$ and the identity operator on $\mathbb R^{2n-4I}_{(n-2I,n-2I)}\oplus \mathbb C^{2(I-1)}$ in the definition of the Norden affinor $(\Delta_I)_\myf\Lambda{}^\myf\Psi=$ $=\frac{1}{2}(\delta_\myf\Lambda{}^\myf\Psi+I_\myf\Lambda{}^\myf\Omega(f_I)_\myf\Omega{}^\myf\Psi)$, where $I_\myf\Lambda{}^\myf\Omega|_{\mathbb R^{2n-4I}_{(n-2I,n-2I)}\oplus \mathbb C^{2(I-1)}}=\delta_\myf\Lambda{}^\myf\Omega|_{\mathbb R^{2n-4I}_{(n-2I,n-2I)}\oplus \mathbb C^{2(I-1)}}$, $I_\myf\Lambda{}^\myf\Omega|_{\mathbb R^4_{(2,2)}}=i\delta_\myf\Lambda{}^\myf\Omega|_{\mathbb R^4_{(2,2)}}$. So it is necessary to answer on the question to justify the spinor decomposition of the Norden affinor: How to construct spinor analogs of complex orthogonal transformations? All significant complex representations are equivalent in all its properties, and therefore $K$ will be omitted if the number of a considered representation is not significant.

\item \textbf{Complex orthogonal transformations. Involutions.}

\hspace{0.5cm} More than once, we will be faced with theorems named after \'{E}lie Cartan in connection with spinors. It is one of them. The theorem was generalized to quadratic forms over arbitrary fields by Dieudonn\'e.
\begin{theorem}(Cartan-Dieudonn\'e)\cite[p. 33(rus)]{Berger1} Let $E$ be a vector space of dimension $n\ge 1$. Every isometry $f \in O(g)$ which is not the identity is the composition of at most $n$ reflections with respect to hyperplanes, where $n=dim\ E$ and g is a nondegenerate quadratic form on $E$.
\end{theorem}
In 1938, Cartan was published this result in \cite[Chapter I, Section 10]{Cartan1}. Thus, any complex orthogonal transformation $S_\Lambda{}^\Psi\ (\Lambda,\Psi,...=\overline{1,n},\ A,B,...=\overline{1,N})$ can be represented as a product of elementary transformations of the form $\pm(r_\Lambda r^\Psi-\delta_\Lambda{}^\Psi)$. The elementary transformations can be decomposed with the help of the triple product for the connecting operators as $(r_\Lambda r^\Psi-\delta_\Lambda{}^\Psi)\eta_\Psi{}^{AB}=$ $=\eta_\Lambda{}_{CD}(r^\Omega\eta_\Omega{}^{CB})(r^\Psi\eta_\Psi{}^{AD})$ \cite[pp. 272-275]{Postnicov1}. Then on the spinor space, spinor analogs from the $Spin(g,n)$ for special orthogonal transformations  are induced \cite[p. 28, eq. (6.42)(eng), p. 156(33), eq. (6.42)(rus)]{Andreev1}, \cite[p. 33, eq. (6.42)]{Andreev0}.
$$
S_\Lambda{}^\Psi\eta_\Psi{}^{AB}=\eta_\Lambda{}^{CD}\tilde S_C{}^A\tilde{\tilde S}_D{}^B\spsd
$$
In the case when the algebra identity preserved, $S_C{}^A:=\tilde S_C{}^A=\tilde{\tilde S}_C{}^A$. For the octonions, the automorphism group keeps the algebra identity, therefore the controlling spin-tensor must be invariant under spinor orthogonal transformations corresponding to automorphism group transformations that reduces the dimension on 7 else. In more detail, the block-diagonal structure can be found in \cite[p. 94-109]{Rosenfeld2}. For the non-special orthogonal transformation the decomposition
$$
S_\Lambda{}^\Psi\eta_\Psi{}^{BA}=\eta_\Lambda{}_{CD}\tilde S^{CA}\tilde{\tilde S}^{DB}
$$
is performed. This makes it possible to pass to infinitesimal transformations by the equation \cite[p. 52, eq. (10.6)(eng), p. 184(61), eq. (10.6)(rus)]{Andreev1}, \cite[p. 61, eq. (10.6)]{Andreev0}.
$$
T_C{}^A=\frac{1}{2}T^{\Theta\Phi}\eta_\Phi{}^{AB}\eta_\Theta{}_{CB}\spsd
$$
For n=8, it is easy to calculate the dimension of the automorphism group for alternative-elastic algebras using this equation \cite{Andreev5}. For the involution, for the given signature, only one of the two decompositions
$$
S_\Lambda{}^{\Psi '}\bar\eta_{\Psi '}{}^{A'B'}=\eta_\Lambda{}^{CD}\tilde S_C{}^{A'}\tilde{\tilde S}_D{}^{B '}\sps
S_\Lambda{}^{\Psi'}\bar\eta_{\Psi'}{}^{B'A'}=\eta_\Lambda{}_{CD}\tilde S^{CA'}\tilde{\tilde S}^{DB'}
$$
is executed \cite[p. 29, eq. (6.47)(eng), p. 156(33), eq. (6.47)(rus)]{Andreev1}, \cite[p. 33, eq. (6.47)]{Andreev0}.

\item \textbf{Klein-Fock-Gordon equation.}

\hspace{0.5cm} In the first time (1926), the Klein-Fock-Gordon equation $(\square -m^2)\phi=0$ was considered by Schrodinger, but was submitted by Vladimir Fock in 1927.  In 1928, British physicist Paul Dirac formulated the equation, the simplified record of which has the form $(i\gamma^i\partial_i-m)\phi=0$ ($i=\overline{1,4}$). In order to get the Klein-Fock-Gordon equation, the Dirac equation is necessary multiplied by $(i\gamma^i\partial_i+m)$ that will lead to the Clifford equation for $\mathbb R^4_{(1,3)}$. This initiated the study of Clifford equation solutions for $n=4$ \cite{Penrose1}. However, it is more convenient to consider the inclusion $\mathbb R^4_{(1,3)}\rightarrow \mathbb C^4$ and to pass to the spinor formalism for the complex space $\mathbb C^4$ and the real representation $\mathbb R^8_{(4,4)}$. Why? Because there is an universal geometric construction: the Cartan triality principle (in 1925) \cite[p. 175(rus)]{Cartan1}, \cite[p. 534]{Rosenfeld2}. In the future, one can always go back with the help of the appropriative real inclusion using the reverse motion. For example, for the inclusion $H_i{}^\Lambda:\mathbb R^4_{(1,3)}\rightarrow \mathbb C^4$, one can define the Infeld-van der Waerden symbols $g_i{}^{AA'}:=H_i{}^\Lambda \eta_\Lambda{}^{AB}S_B{}^{A'}$ ($i,j,...=\overline{0,3}$, $\Lambda=\overline{0,3}$, $A,B,A',B'...=\overline{1,2}$). By this equation, the involution $S_B{}^{A '}$ is introduced in the definition of the symbols $g_i{}^ {AA'}$. Accordingly, the involution $S_A{}^{A'}$ will generate the involution $\bar S$ for the complex Dirac operators $\bar\gamma_i = \bar S\gamma_i\bar S$, and in this special basis, $||S||=||\gamma^2||$ will be executed. Thus, we can define the Majorana spinor as
$\psi^i=\gamma^2 \bar\psi^i$ which corresponds to the presentation \cite[p. 98, Appendix E (eng)]{Zee1}, \cite[p. 89, Section 14.7.1(eng), p. 225(102), Section 14.5.6(rus)]{Andreev1}, \cite[p. 102, Section 14.5.6]{Andreev0}.

\item \textbf{Cartan triality principle.}

\begin{theorem}
\label{theorem1}(The triality principle for two B-cylinders)\cite[p. 91, Section 14.7.2(eng), p. 227(104), Section 14.5.7(rus)]{Andreev1},
\cite[p. 104, Section 14.5.7]{Andreev0}, \cite[p. 70, Section 5.7(eng), p. 183(94), Section 5.7(rus)]{Andreev2}, \cite{Andreev6}. In the projective space $\mathbb C \mathbb P_7$, there are two quadrics (two B-cylinders) with the following main properties:
\begin{enumerate}[1.]
    \item The planar generator $\mathbb C \mathbb P_3$ of a one quadric will define one-to-one the point R on the other quadric.
    \item The planar generator $\mathbb C \mathbb P_2$ of a one quadric will uniquely define the point R on the other quadric. But the point R of the second quadric can be associated to the manifold of planar generators $\mathbb C \mathbb P_2$ belonging to the same planar generator $\mathbb C \mathbb P_3$ of the first quadric.
    \item The rectilinear generator $\mathbb C \mathbb P_1$ of a one quadric will define one-to-one the rectilinear generator $\mathbb C \mathbb P_1$ of the other quadric. And all the rectilinear generators belonging to the same planar generator $\mathbb C \mathbb P_3$ of the first quadric define the beam centered at R belonging to the second quadric.
\end{enumerate}
\end{theorem}
\begin{enumerate}[$\bullet$]
\item For even n $<$ 8 the dimension of the base space is greater than the one of the spinor space.
\item For n = 8 the dimension of the base space is equal to the one of the spinor space.
\item For even n $>$ 8 the dimension of the base space is less than the one of the spinor space.
\end{enumerate}
Thus, for n $\le$ 8, one can construct the division normalized algebra. For n $>$ 8, the 0-divisors appear, because the inclusion of the base space in the spinor one is performed with the help of the inclusion operator $P^i{}_B:=\eta^i{}_{AB}X^A$, which is included in the definition of algebraic basis elements and ultimately determines the 0-divisors of an algebra itself. The weakly normalization identity has the form $g_{kr}\eta_{\left(\right.i|j|}{}^k\eta_{l\left.\right)m}{}^r=
g_{kr}\eta^k{}_{j\left(\right.i}\eta^r{}_{|m|l\left.\right)}$. This identity is the normalization identity for n=8 only ($g_{kr}\eta^k{}_{j\left(\right.i}\eta^r{}_{|m|l\left.\right)}=g_{jm}g_{il}$ in this case). What is based the triality principle on?

\item \textbf{Rosenfeld null-pairs.}

\hspace{0.5cm} Rosenfeld null-pairs lie in the heart of the triality principle. Let the pair $X^A:=(X^b,Y_a)\ (a,b,...=\overline{1,4}\sps A,B,...=\overline{1,8})$ be \emph{Rosenfeld null-pair} \cite{Rosenfeld1}, \cite[p. 378]{Rosenfeld3}. In the space $\mathbb C\mathbb P^4=~'\mathbb C^4/'\mathbb C$ (where $'\mathbb C^n=\mathbb C^n/{0}$), $X^b$ defines the point and $Y_a$ defines the plane with \emph{incidence condition} $X^b Y_a=0$ \cite[p. 362]{Rosenfeld3}. Therefore, we can construct the space $\mathbb C\mbox{\foreignlanguage{russian}{Ï}}^4='\mathbb C^*{}^4/'\mathbb C$, which is \emph{dual space} to $\mathbb C\mathbb P^4$. Then the space $\mathbb C\mathbb P^4\times \mathbb C\mbox{\foreignlanguage{russian}{Ï}}^4$ is \emph{Rosenfeld null-pair space}. It should be noted that such the spaces were studied by Sintcov \cite{Sintcov1} and  Kotelnikov \cite{Kotelnikov1} for the first time. Both a twistor is constructed form pair spinors and a bitwistor consists of two twistors, and an isotropic bitwistor is the null-pair with the incidence condition. The isotropic bitwistor $X^A$ is associated to the parity $X^a=ir^{ab}Y_b,\ r^{ab}=-r^{ba}:=r^\alpha\eta_\alpha{}^{ab}\ (\alpha,...=\overline{1,6})$ as well as an twistor is associated to pair spinors \cite[v. 2, p. 47, eq. (6.1.13)]{Penrose1}, where $r^{ab}$ is the same as in \cite[v. 2, p. 306, eq. (9.3.7)]{Penrose1} for the real inclusion in the special basis.
A vector of the base space is determined by analogy with \cite[v. 2, p. 306-307]{Penrose1} just as \cite[p. 83-84(eng), p. 218-223(95-100)(rus)]{Andreev1}.
The metric tensor $g_{\Lambda\Psi}\ (\Lambda,\Psi,...=\overline{1,8})$ acts on the base space, and its metric analog $\varepsilon_{AB}$ acts on the bitwistor space.  Thus, the two quadrics are define:
\begin{enumerate}[$\bullet$]
\item the quadric, determined with the help of $\varepsilon_{AB}X^AX^B=0$;
\item the quadric, determined by the 6 parameters $r^{ab}$ with the help of $g_{\Lambda\Psi}r^\Lambda r^\Psi~=~0$.
\end{enumerate}
This quadrics are the B-cylinders from Theorem \ref{theorem1}. This allows to obtain the representation of an isotropic twistor on the isotropic cone of the base space $R^6_{(2,4)}$. This is completely analogous to the representation of a spinor on the isotropic cone of the space-time with only the one difference: the dimensions of the flag and the flagpole are increased by 1 \cite[p. 72-77(eng), p. 205-210(82-87)(rus)]{Andreev1}. In order to construct the representation, one can need to use the result of the paper \cite{Rozenfeld4} and to define the matrix differential. Besides, the Theorem \ref{theorem1} combines both the Cartan triality principle and the Klein correspondence \cite[v. 2, p. 307-313]{Penrose1}.

\item \textbf{Spinor formalism. Partial solutions of the Clifford equation.}

\hspace{0.5cm} Solutions of the Clifford equation for n=4 are given in \cite{Penrose1} and associated to the Pauli matrixes. As is known, they are 4: an one antisymmetric matrix and three symmetric matrixes. In Russia, for n=6, the first appearance of partial solutions of the Clifford equation seems to be attributed to \cite{Norden1} in 1959. They are related to Plucker coordinates for the Grassmannian $G(2,4)$. To foreign publications, the article \cite{Klotz1},\cite{Buchdahl1} can be attributed. Further development of these results were obtained by Ukrainian physicist Yu. P. Stepanpovski\u\i\ \cite{Stepanovskii1}, \cite{Stepanovskii2}. Thus, for n=6, the partial solutions of the Clifford equation can be represented with the help of six antisymmetric matrixes. It should be noted that the specified formalism is closely connected with the Bogolyubov-Valatin transformations \cite{Scharnhorst1}. It can be shown that for n=8, partial solutions of the Clifford equation can be represented with the help of 8 matrixes: seven antisymmetric matrixes and an one symmetric matrix \cite{Andreev3}, \cite[p. 77, Section 14.4(eng), p. 211(88), Section 14.4 (rus)]{Andreev1}, \cite[p. 88, Section 14.4]{Andreev0}. Therefore, we must assume that for even n, partial solutions of the Clifford equation can be represented with the help of $n$ matrixes: $\tilde q$ antisymmetric matrixes and $q$ symmetric matrixes $(\tilde q+q=n)$. Note, that the Penrose spinor formalism for n=4 is based on the Klein correspondence \cite{Penrose1}, while the spinor formalism for even n can be considered on the base of the Cartan triality principle \cite{Andreev7}, \cite{Andreev0}-\cite{Andreev2}. And because the Klein correspondence is a particular case of the Cartan triality principle then it makes sense to consider the spinor formalism for n=8 and then to go back to n=4 with the help of the reverse motion.

\item \textbf{Bott periodicity. Classification of metric group alternative-elastic algebras.}

\hspace{0.5cm} In 1908, Cartan discovered the principle of periodicity for the Clifford algebras \cite{Cartan2}, \cite{Baez1}, \cite{Postnicov1} $CL(g^{n+8}_{(n+8,0)})\cong CL(g^n_{(n,0)})\otimes\mathbb R[16]$. But this property is true for any signature of the metric. This means that the main properties must be preserved for algebras with the same value n mod 8. The presence of an symmetric metric spinor on the spinor space provides the existence of the group structure with respect to multiplication for n mod 8 = 0. This implies that since octonions are endowed with the group alternative-elastic structure then alternative-elastic algebras exist for any n mod 8=0. Moreover, octonions is initial induction step for the construction of such the algebras for n mod 8 = 0. To construct the next even-dimensional connecting operators, one can use the inductive transition of the form
$$
\begin{array}{lc}
 &
\begin{array}{c}
\eta_\Lambda{}^{AB}=
\left(
\begin{array}{cc}
 \eta_\alpha{}^{ab}                            &  \frac{1}{2}(i\eta_{n-1}+\eta_n)\delta^a{}_d  \\
\frac{1}{2}(-i\eta_{n-1}+\eta_n)\delta_c{}^b   &                      -(\eta^T)_\alpha{}_{cd}
\end{array}
\right)\sps\\
\sigma_\Lambda{}_{AB}=
\left(
\begin{array}{cc}
           (\eta^T)_\alpha{}_{ab}              &   \frac{1}{2}(i\eta_{n-1}+\eta_n)\delta_a{}^d \\
 \frac{1}{2}(-i\eta_{n-1}+\eta_n)\delta^c{}_b  &                           -\eta_\alpha{}^{cd}
\end{array}
\right)
\end{array}
\end{array}
$$
with different variations of this form as shown in Algorithm 9.1 \cite[p. 50(eng), p. 181(58)(rus)]{Andreev1}, \cite[p. 58]{Andreev0}, in Algorithm 1 \cite{Andreev4}; $\Lambda=\overline{1,n}$; $\alpha=\overline{1,n-2}$; $a,b,...=\overline{1,N/2}$; $A,B,C,D,K=\overline{1,N}$; $N:=2^{n/2-1}$. Note, that similar constructions are considered in \cite[p. 518]{Rosenfeld2} for alternions. Besides, such the transition is given in \cite[p. 65, eq. (6.2.18)]{Penrose1}. Thus, the connecting operation can be constructed inductively for any n mod 8=0. For inclusion $H_i{}^\Lambda$: $\mathbb R^n\rightarrow \mathbb C^n,\ (i,j,...=\overline{1,n})$, any metric alternative-elastic algebra can be expanded into the basis, the I-element of which is the algebra of the form
$$
(\eta_I)_{ij}{}^k:=\sqrt{2}\eta_i{}^{AB}\eta_{j CA}\eta^k{}_{DB}((X_I)^C(X_I)^D+(1-\frac{1}{2}(X_I)^K(X_I)_K)\frac{2}{N}\varepsilon^{CD})\sps
$$
where $\varepsilon^{CD}$ is the metric spinor on the spinor space, and $(1-\frac{1}{2}(X_I)^K(X_I)_K)\frac{2}{N}\varepsilon^{CD}$ guarantee the existence of the algebra identity. We will consider such  the orthogonal basis elements $(X_I)^C$, which are transformed into each other with the help of the orthogonal transformation only and $(X_I)^C(X_I)_C:=2$. Indeed, any controlling symmetric spinor $\theta^{CD}$ can be decomposed as
$$
\theta^{CD}=\sum\limits_{I=1}^J\alpha_I\underbrace{(X_I)^C(X_I)^D}_{:=(\theta_I)^{CD}}+ \underbrace{(1-\sum\limits_{I=1}^J\alpha_I)}_{:=\alpha_0}\underbrace{\frac{2}{N}\varepsilon^{CD}}_{(\theta_0)^{CD}}\spsd
$$
In particular, Cayley-Dickson algebras admit such the decomposition and are special orthogonal metric alternative-elastic algebras. This means, that such the algebra is generated by special orthogonal transformations from the single basis element. This element is generated from the controlling spinor with the significant coordinates $X^1:=1\sps X^{2^{\frac{n}{2}-2}+1}:=1$ \cite[p. 8]{Andreev4}. The construction of the sedenion algebra, based on this expansion, is realized in Appendix \cite{Andreev4}. One can easy classify these algebras for n = 8. This classification is based on the one of eigenvalues of the controlling spin-tensor $\theta^{CD}$ \cite{Andreev5}. For the octonion algebra \cite[ex. 1]{Andreev5}, $\theta^{CD}$ has the single significant eigenvalue, and therefore the dimension is reduced on 7 else. A similar classification can be constructed for any n mod 8 = 0.

\item \textbf{Complex and real representations of (pseudo-)Riemannian spaces.}

\hspace{0.5cm} Under \emph{complex analytical Riemannian space} $\mathbb CV_n$, we will further understand an analytical complex manifold supplied with \emph{analytical quadratic metric}, i.e., a metric defined by means of a symmetric nonsingular tensor $g_{\Lambda\Psi}$ (here $\Lambda , \Psi ,... = \overline {1, n} $; $\myf\Lambda ,\myf\Psi ,... = \overline {1,2n}$), which the coordinates are analytical functions of the point coordinates. To this tensor, there corresponds \emph{complex Riemannian torsion-free connection}, the coefficients of which are defined by the Christoffel symbols, and hence these coefficients are analytical functions. The tangent bundle to this manifold $\tau^\mathbb C (\mathbb CV_n)$ has fibers $\tau_x^\mathbb C\cong \mathbb C\mathbb R^n $ that is fibers isomorphic to the n-dimensional complex Euclidian space, the metric of which is defined by the value of the metric tensor at the given point. The real representation $V_{2n}$ of $\mathbb CV_n$ has the tangent bundle $\tau^\mathbb R(V_{2n})$ with fibers isomorphic to $\mathbb R^{2n}_{(n, n)}$. Let an atlas $(U; x^\myf\Lambda)$ be set on $V_{2n}$. We will consider \emph{reparametrization} of this atlas $(U; w^\Lambda)$ such that $w^\Lambda=\frac{1}{\sqrt{2}}(u^\Lambda(x^\myf\Lambda)+iv^\Lambda (x^\myf\Lambda))$, which is locally solvable as $x^\myf\Lambda= x^\myf\Lambda (u^\Lambda, v^\Lambda)$. Set
$$
 m^\Lambda{}_\myf\Lambda:=\frac{1}{\sqrt{2}}(
 \frac{\partial u^\Lambda}{\partial x^\myf\Lambda}+i
 \frac{\partial v^\Lambda}{\partial x^\myf\Lambda})=:\frac{\partial w^\Lambda}{\partial x^\myf\Lambda}\sps
 m_\Lambda{}^\myf\Lambda:=\frac{1}{\sqrt{2}}(
 \frac{\partial x^\myf\Lambda}{\partial u^\Lambda}-i
 \frac{\partial x^\myf\Lambda}{\partial v^\Lambda})=:\frac{\partial x^\myf\Lambda}{\partial w^\Lambda}\sps
$$
$$
f_\myf\Lambda{}^\myf\Psi:=\frac{\partial v^\Lambda}{\partial x^\myf\Lambda}\ \frac{\partial x^\myf\Psi}{\partial u^\Lambda}-
   \frac{\partial u^\Lambda}{\partial x^\myf\Lambda}\ \frac{\partial x^\myf\Psi}{\partial v^\Lambda}\spsd
$$
Then the operators $m^\Lambda{}_\myf\Lambda$ are coincide with the ones from Section 8 \cite[p. 9, Section 3(eng), p. 133(10), Section 3(rus)]{Andreev1}, \cite[p. 10, Section 3]{Andreev0}.

\item \textbf{Complex and real connections over Riemannian spaces. Case I.}

\hspace{0.5cm}
Therefore, if one can define the connection $\nabla_\myf\Phi G_{\myf\Lambda\myf\Psi}=0$ ($\myf\Lambda,\myf\Psi ,... = \overline {1,2n}$) in the tangent bundle for the real representation then the connection $\nabla_\Phi g_{\Lambda\Psi}=$ $=0\sps \bar\nabla_{\Phi '} \bar g_{\Lambda'\Psi'}=0$ ($\Lambda,\Psi ,... = \overline {1,n}$) is induced in the tangent bundle for the complex representation by the covariant constancy of the Neifeld operators. Then onto the spinor bundle with fibers isomorphic $\mathbb C^N$, the connection prolongs with the help of the covariant constancy of the connecting operators. If someone can want to move into the real inclusion then he must request the covariant constancy of the inclusion operator that will lead to the Riemannian connection $\nabla_i g_{jk}=0$ ($i,j,... = \overline {1,n}$) \cite[p. 59, Corollary 11.1(eng), p. 190(67), Corollary 11.1(rus)]{Andreev1}, \cite[p. 67, Corollary 11.1]{Andreev0},\cite[Section 3]{Andreev2}. In the last case, on the spinor bundle, the induced involution must be the covariant constant. On the spinor space, such the connection is called \emph{connection compatible with involution}. Hereinafter, all connections are torsion-free.

\item \textbf{Revers motion: Lie operator analogues. Spin-pair space. Case II.}

\hspace{0.5cm} The connection can be prolonged an another way onto the spinor bundle \cite[p.59, Theorem 11.2(eng), p. 191(68), Theorem 11.2(rus)]{Andreev1}, \cite[p. 68, Theorem 11.2]{Andreev0}. On a tangent fiber of the real representation $V_{2n}$, one can define the operator $P^\myf\Psi{}_{\myff A}:=\eta^\myf\Psi{}_{\myff B\myff A}X^{\myff B}$ and $P_\myf\Psi{}^{\myff A}:=\eta_\myf\Psi{}^{\myff B\myff A}Y_{\myff B}$ ($\myf\Lambda ,\myf\Psi ,... = \overline {1,2n}$, $\myff A ,\myff B ,... = \overline {1,2N^2}$) such that $X^{\myff A}Y_{\myff A}=2$ and $P_\myf\Lambda{}^{\myff A}P_\myf\Psi{}_{\myff A}=G_{\myf\Lambda\myf\Psi}$. This is always possible for $n\ge 8$. Then the compatibility condition of the connections for the real representation will have the form $\nabla_\myf\Lambda P_\myf\Psi{}^{\myff A}=0$. For the complex representation, the operators
$(P_K)^\Psi{}_A:=m_\myf\Psi{}^\Psi P^\myf\Psi{}_{\myff A}(\tilde M_K)_A{}^{\myff A}\sps$
$(P_K)^{\Psi '}{}_A:=\bar m_\myf\Psi{}^{\Psi '} P^\myf\Psi{}_{\myff A}(\tilde M_K)_A{}^{\myff A}\sps$
$(P^*_K)_\Psi{}^A:=m^\myf\Psi{}_\Psi P_\myf\Psi{}^{\myff A}(\tilde M^*_K)^A{}_{\myff A}\sps$
$(P^*_K)_{\Psi '}{}^A:=\bar m^\myf\Psi{}_{\Psi '} P_\myf\Psi{}^{\myff A}(\tilde M^*_K)^A{}_{\myff A}$ ($\Lambda,\Psi ,... = \overline {1,n}$, $A ,B = \overline {1,N}$, $K = \overline {1,2N}$)
are defined. Then for the K-th complex representation, the Riemannian connection determines by the covariant constancy of the Neifeld operators and the operators $\myf M_K, \myf M_K^*$. This induces the prolongation of the connection onto the complex spinor bundle by the covariant constancy of the operators $P_K,P^*_K$. Analogically, for the real inclusion, it is necessary to demand the covariant constancy of the inclusion operator $H$. If someone demands not only the covariant constancy but also the ordinary one then he can define the Lie operator analogues on the spinor bundle \cite[p.61, Theorem 11.3(eng), p. 192(69), Theorem 11.3(rus)]{Andreev1}, \cite[p. 69, Theorem 11.3]{Andreev0}
$$
\begin{array}{l}
L_x(Y_K)^A:=x^\Omega\partial_\Omega (Y_K)^A-\sum\limits_{J=1}^{2N}(Y_J)^B(P_J)^\Psi{}_B(P^*_K)_\Omega{}^A\partial_\Psi x^\Omega\sps\\
\bar L_{\bar x}(\bar Y_K)^A:=\bar x^{\Omega '}\bar\partial_{\Omega '}(\bar Y_K)^A-\sum\limits_{J=1}^{2N}(\bar Y_J)^B(\bar P_J)^{\Psi '}{}_B(\bar P^*_K)_{\Omega '}{}^A\bar\partial_{\Psi '}\bar x^{\Omega '}\spsd\\
\end{array}
$$
For the inclusion $\mathbb R^4_{(1,3)}\rightarrow \mathbb C^4$ (n=4), the constructed operators have the form \cite[p.96, Example 14.2(eng), p. 262, Addition (rus)]{Andreev1}
$$
\begin{array}{c}
(P^*_K)_i{}^A:=
\frac{1+i}{4}\left(
\begin{array}{cc}
 0          &  p_1-p_2   \\
 0          &  1-p_1p_2  \\
 p_1-p_2    &  0         \\
 i(1-p_1p_2)&  0         \\
\end{array}
\right)\sps\\ \\
(P_K)^i{}_A:=
\frac{1-i}{4}\left(
\begin{array}{cc}
  0          &  p_1-p_2   \\
  0          &  1-p_1p_2  \\
  p_1-p_2    &  0         \\
-i(1-p_1p_2) &  0         \\
\end{array}
\right)\sps
\end{array}
$$
where $p_1:=\pm 1$, $p_2:=\pm 1$. Therefore, one can construct the one-to-one mapping ($a,b,...=\overline{1,4}$, $A,B,...=\overline{1,2}$, $i,j,...=\overline{1,4}$)
$$
x^i\cdot\underbrace{((P^*_2)_i{}^A,(P^*_3)_i{}^B)}_{:=P_i{}^a}=\underbrace{((X_2){}^A,(X_3){}^B)}_{:=X^a}\sps
x^i\cdot\underbrace{((P_2)_i{}_A,(P_3)_i{}_B)}_{:=P_i{}_a}=\underbrace{((X^*_2){}_A,(X^*_3){}_B)}_{:=X_a}\spsd
$$
Then on the spin-pair $X^a=((X_2){}^A,(X_3){}^B)$, the metric and the involution of the form
$$
\begin{array}{c}
\varepsilon_{ab}=
\left(
\begin{array}{cc}
                      0 & (P_2)^i{}_A(P_3)_i{}_B \\
 (P_3)^i{}_C(P_2)_i{}_D & 0 \\
\end{array}
\right)=
\left(
\begin{array}{cccc}
 0 & 0 &-i & 0\\
 0 & 0 & 0 & i\\
-i & 0 & 0 & 0\\
 0 & i & 0 & 0\\
\end{array}
\right)\sps\\
S_a{}^{b'}=
\left(
\begin{array}{cc}
 (P_2)^i{}_A(\bar P^*_2)_i{}^{C'} & (P_2)^i{}_A(\bar P^*_3)_i{}^{D'} \\
 (P_3)^i{}_B(\bar P^*_2)_i{}^{C'} & (P_3)^i{}_A(\bar P^*_3)_i{}^{D'} \\
\end{array}
\right)=
\left(
\begin{array}{cccc}
 0 & 0 & i & 0\\
 0 &-i & 0 & 0\\
 i & 0 & 0 & 0\\
 0 & 0 & 0 &-i\\
\end{array}
\right)
\end{array}
$$
are defined. Therefor, it is possible to construct \emph{Lie pair-spin operators}
$$
L_x(Y)^a=x^i\nabla_iY^a-P^j{}_bY^bP_i{}^a\nabla_j x^i\sps
L_X(Y)^a=X^c\nabla_cY^a-Y^c\nabla_c X^a\spsd
$$
Note, that the expansion $x^i:=(P_2)^i{}_A(X_2){}^A+(P_3)^i{}_A(X_3){}^A$ is an one of the vector into the two isotropic vectors. And for the isotropic vectors, \emph{Lie spinor operators} were constructed in \cite[v. 2, pp. 101-103 (eng)]{Penrose1}. The restricted Lorentz transformation converts the vector $x^i$ to the vector $\tilde x^i$. This induces the spinor transformation: $(X_K)^A\longmapsto (\tilde X_K)^A$. The spinor transformations act on the operators $(P^*_K)_i{}^A,\ (P_K)^i{}_A$, more precisely, on the controlling spinor $X^\myff A$ appearing in the definition of these operators. In turn, this induces the spin-pair transformation of the spin-pair $X^a=((X_2){}^A,(X_3){}^B)\longmapsto \tilde X^a$.

\item \textbf{Normalization of Grassmannians. Twistor equation. Case III.}

\hspace{0.5cm} \emph{Normalization} of the complex Grassmannian $G_\mathbb C(2N,4N)$ \cite{Newfield2} is an analytical differential map $N$ of a domain $D\subset G_\mathbb C(2N,4N)$ into the dual Grassmannian $G^*_\mathbb C(2N,4N)$. $(\mathbb C^*)^{2N}(Y_\myf A)$ ($\myf A,\myf B,...=\overline{1,2N}$) is called \emph{normalizative subspace for subspace} $\mathbb C^{2N}(X_\myf B)$.  The last must not have common directions with the first. $Y_\myf A$, $X_\myf B$ are basis elements of the corresponding subspace. In local coordinates, the correspondence $N$ must be given with the help of the parametric equations $X_\myf B=X_\myf B(r^\Lambda)$, $Y_\myf A=Y_\myf A(r^\Lambda)$ ($\Lambda,\Psi,...=\overline{1,n}$). The boundary points of $D$, if they exist and for which the plane $\mathbb C^{2N}$ has an one common direction with $(\mathbb C^*)^{2N}$ at least, form \emph{absolute of the normalization}. The derivational equations  of the normalized Grassmannian can have the form
$\nabla_\Lambda X_\myf A=i\gamma_{\Lambda \myf A}{}^\myf B Y_\myf B$, $\nabla_\Lambda Y_\myf B=0$ in the particular case. If $\gamma_\Lambda$ are satisfied the Clifford equation then such the normalization is called \emph{spinor normalization}, and the derivational equations \cite[eq. (1.2)]{Newfield2} can be reduced to $\nabla_\Lambda X^A=\underbrace{i\eta_\Lambda{}^{AB} Y_B}_{:=P_\Lambda{}^A}, Y_B=const$ ($A,B,...=\overline{1,N}$). They can be rewritten as the twistor equation \cite[v. 2, p. 463, eq. (B.94a)]{Penrose1}
$$
\eta_\Lambda{}_{AB}\nabla_\Psi X^A+\eta_\Psi{}_{AB}\nabla_\Lambda X^A=\frac{2}{n}g_{\Lambda\Psi}\eta^\Phi{}_{AB}\nabla_\Phi X^A\spsd
$$
For any $X^A$, the integrability condition is write down as $C_{\Lambda\Psi\Phi\Omega}=0$ \cite[Section 13]{Andreev1}, where $C_{\Lambda\Psi\Phi\Omega}$ is the Weyl tensor. Thus, the constructed connection is conformally Euclidean. For this case, the solution of the twistor equation  has the form $X^C:=\dot X^C+iR^{CA}\dot Y_A$, $\dot X^C=const$, $\dot Y_A=const$, $R^{CA}:=r^\Lambda\eta_\Lambda{}^{CA}$. Recall that there are the $2N$ units of such the complex representations. If someone will find the sum on $K$ then he will obtain the Killing equation from the twistor equation \cite[Section 13, p. 68(eng), p. 200(77)(rus)]{Andreev1}. For n=6, the particular case  is considered in \cite{Andreev8}.

\end{enumerate}

}
\References{
\bibitem{Albert1}
A.A. Albert. Quadratic Forms Permitting Composition. Ann. of Math. 1942, 43, 161-177.
\bibitem{Andreev0}
К.В. Андреев  [K.V. Andreev]. О спинорном формализме при четной размерности базового пространства [O spinornom formalizme pri chetno\u\i razmernosti basovogo prostranstva]. ВИНИТИ - 298-B-11 [VINITI-298-V-11], июнь 2011 [Jun 2011], 138 с [138pp].  [in Russian: On the spinor formalism for the base space of even dimension. Paper deponed on Jun 16, 2011 at VINITI (Moscow), ref. \No 298-V 11]
\bibitem{Andreev1}
K.V. Andreev. On the spinor formalism for even n. \href{http://arxiv.org/abs/1110.4737}{arXiv:1110.4737v3} [math-ph].
[with the Russian edition: К.В. Андреев. О спинорном формализме при четной размерности базового пространства.].
\bibitem{Andreev2}
К. В. Андреев [K.V. Andreev]: Спинорный формализм и геометрия шестимерных римановых пространств [Spinorny\u\i\ formalizm i geometriya shestimernykh rimanovykh prostranstv]. Кандидатская диссертация [Kandidatskaya dissertatsiya], Уфа [Ufa], 1997, [in Russian: Spinor formalism and the geometry of six-dimensional Riemannian spaces. Ph. D. Tesis], \href{http://arxiv.org/abs/1204.0194}{arXiv:1204.0194v1} (with the Russian edition).
\bibitem{Andreev3}
К. В. Андреев [K.V. Andreev]: О структурных константах алгебры октав. Уравнение Клиффорда [O strukturnykh konstantakh algebry oktav. Uravnenie Klifforda]. Изв. Высш. Учебн. Завед., Матем. [Izv. Vyssh. Uchebn. Zaved. Mat.],  3(2001)3-6. English translation: K. V. Andreyev: Structure constants of the algebra of octaves. The Clifford equation. Russian Mathematics (Iz. VUZ), 45:3(2001)1-4 (Mathnet URL: \url{http://mi.mathnet.ru/ivm856}).
\bibitem{Andreev4}
K.V. Andreev. On the metric hypercomplex group alternative-elastic algebras for n mod 8 = 0. \href{http://arxiv.org/abs/1202.0941}{arXiv:1202.0941v1} [math-ph]
\bibitem{Andreev5}
K.V. Andreev. On the classification of metric hypercomplex group alternative-elastic algebras for n=8. \href{http://arxiv.org/abs/1208.4466}{arXiv:1208.4466v1} [math-ph]
\bibitem{Andreev6}
К. В. Андреев [K.V. Andreev]: Принцип тройственности для двух квадрик [Printsip tro\u\i stvennosti dlya dvykh kvadrik]. ВИНИТИ-2470-B-98 [VINITI-2470-V-98], август [Aug] 1998. 19 с. [19pp. Paper deponed on Aug 3, 1998 at VINITI (Moscow), ref. \No 2470-V 98.], [in Russian: Triality principle for two quadrics].
\bibitem{Andreev7}
К. В. Андреев [K.V. Andreev]: О твисторах 6-мерного пространства [O tvistorakh 6-mernogo prostranstva]. ВИНИТИ-2469-B-98 [VINITI-2469-V-98], август [Aug] 1998, 23 с [23pp. Paper deponed on Aug 3, 1998 at VINITI (Moscow), ref. \No 2469-V 98], [in Russian: On twistors of 6-dimensional space].
\bibitem{Andreev8}
К. В. Андреев [K.V. Andreev]: О внутренних геометриях многообразия плоских образующих 6-мерной квадрики [O vnutrennikh geometriyakh mnogoobraziya ploskikh obrazuyushchikh 6-merno\u\i\ kvadriki], Изв. Высш. Учебн. Завед., Матем. [Izv. Vyssh. Uchebn. Zaved. Mat.], 6(1998)3-8. English translation: K. V. Andreyev: On intrinsic geometries of the manifold of plane generators of a 6-dimensional quadric, Izv. Vyssh. Uchebn. Zaved. Mat., 42:6(1998)1-6 (Mathnet URL: \url{http://mi.mathnet.ru/ivm436}).
\bibitem{Baez1}
John C. Baez The Octonions. Bull. Amer. Math. Soc. 39 (2002), 145-205, \href{http://arxiv.org/pdf/math/0105155v4.pdf}{arXiv:math.RA/0105155v4}.
[Баэз Джон С. Октонионы.// Гиперкомплексные числа в геометрии и физике. \No 1(5), Vol. 3 (2006), c.120-177].
\bibitem{Berger1}
Berger, Marcel: Geometry. II. Translated from the French by M. Cole and S. Levy. Universitext. Springer-Verlag, Berlin, 1987. Berger, Marcel: Geometry. I. Translated from the French by M. Cole and S. Levy. Universitext. Springer-Verlag, Berlin, 1987. Russian translation: М. Бергер [M. Berger]. Геометрия [Geometriya], т. 1, 2 [v. 1, 2]. Мир [Mir], Москва [Moskva],  1984. Russian translation by Ю.Н. Сударев [Yu.N. Sudarev], А.В. Пажитнов [A.V. Pazhitnov] and С.В. Чмутов [S.V. Chmutov] under edition И.Х. Сабитов [I.Kh. Sabitov].
\bibitem{Buchdahl1}
H.A. Buchdahl: On the calculus of for-spinors. Proceedings of the Royal Society of London, Series A, Mathematical and Physical Sciences, 303(1968)355-378 (\href{http://dx.doi.org/10.1098/rspa.1968.0055}{DOI:10.1098/rspa.1968.0055})
\bibitem{Cayley1}
Arthur Cayley, On Jacobi's elliptic functions, in reply to the Rev. B. Bronwin; and on quaternions,
Philos. Mag. 26(1845)208-211
\bibitem{Cartan1}
\'{E}. Cartan: Le\c{c}ons sur la Th\'{e}orie des Spineurs, 2 Vols.. Vol.\ I: Les Spineurs de l'Espace a Trois Dimensions. Actual. Sci. Ind., Vol.\ 643, Expos\'{e}s G\'{e}om., Vol.\ 9. Vol.\ II: Les Spineurs de l'Espace a n $>$ 3 dimensions. Les Spineurs en G\'{e}om\'{e}trie Riemanienne. Actual. Sci. Ind., Vol.\ 701, Expos\'{e}s G\'{e}om., Vol.\ 11. Hermann, Paris, 1938. English translation: The Theory of Spinors. Hermann, Paris, 1966. Reprinted: Dover Publications, Inc., New York, 1981. Russian translation: Э. Картан [\`E Kartan]: Теория спиноров [Teoriya spinorov]. Платон [Platon], Москва [Moskva], 1997. Russian translation by П. А. Широков [P.A. Shirokov].
\bibitem{Cartan2}
\'{E}lie Cartan: Nombres complexes, in Encyclop\'{e}die des sciences math\'{e}matiques, 1, ed. J. Molk, 1908, 329-468.
\bibitem{Hamilton1}
Hamilton William Rowan. On quaternions, or on a new system of imaginaries in algebra. Philosophical Magazine, 25:3(1844)489-495. 1844.
\bibitem{Klotz1}
F. S. Klotz: Twistors and the conformal group. Journal of Mathematical Phisics, 15(1974)2242-2247, (\href{http://dx.doi.org/10.1063/1.1666606}{DOI:10.1063/1.1666606}).
\bibitem{Moreno1}
R. Guillermo Moreno. The zero divisors of the Cayley-Dickson algebras over the real numbers. Sociedad Matematica Mexicana. Boleti n. Tercera Serie 4:1(1998)13-28 \href{http://arxiv.org/pdf/q-alg/9710013}{arXiv:q-alg/9710013v1} [math.QA]
\bibitem{Kotelnikov1}
А.П. Котельников [A.P. Kotel'nikov]: Винтовое счисление и некоторые приложения его к геометрии и механике [Vintovoe schislenie i nekotorye prilozheniya ego k geometrii i mekhanike]. Типо-литография Императорского Университета [Tipo-litografiya Imperatorskogo Universiteta], Казань [Kazan'], 1895. [in Russian: Screw Calculus and Some of Its Applications to Geometry and Mechanics]. Second reprint: А.П. Котельников [A.P. Kotel'nikov]: Винтовое счисление и некоторые приложения его к геометрии и механике [Vintovoe schislenie i nekotorye prilozheniya ego k geometrii i mekhanike]. КомКнига [Komkniga], Москва [Moskva], 2006.
\bibitem{Postnicov1}
М.М. Постников. [M.M. Postnikov] Группы и алгебры Ли [Gruppy i algebry Li]. Наука [Nauka], Москва [Moskva], 1986. English translation: M. Postnikov: Lie Groups and Lie Algebras. Lectures in Geometry, Semester 5. Mir, Moscow, 1986; URSS Publishing, Moscow, 1994. The main ideas of the hypercomplex number constraction on the base of the Bott periodicity are given in the lectures 13-16.
\bibitem{Newfield1}
Э.Г. Нейфельд [\`E.G. Ne\u\i fel'd]: Об инволюциях в комплексных пространствах [Ob involyutsiyakh v kompleksnykh prostranstvakh]. Тр. Геом. Семин. [Tr. Geom. Semin.], Казанский университет [Kazanski\u\i\ universitet] (Выпуск [Vypusk]) 19(1989)71-82 (Mathnet URL: \url{http://mi.mathnet.ru/eng/kutgs98}). [in Russian: Involutions in complex spaces].
\bibitem{Newfield2}
Э.Г. Нейфельд [\`E.G. Ne\u\i fel'd]: Нормализация комплексных грассманианов и квадрик [Normalizatsiya kompleksnykh grassmanianov i kvadrik]. Тр. Геом. Семин. [Tr. Geom. Semin.], Казанский университет [Kazanski\u\i\ universitet] (Выпуск [Vypusk]) 20(1990)58-69 (Mathnet URL: \url{http://mi.mathnet.ru/kutgs82}). [in Russian: Normalization of complex Grassmannians and quadrics].
\bibitem{Norden1}
А.П. Норден [A.P. Norden]: О комплексном представлении тензоров пространства Лоренца [O kompleksnom predstavlenii tenzorov prostranstva Lorentsa]. Изв. Высш. Учебн. Завед., Матем. [Izv. Vyssh. Uchebn. Zaved., Mat.] (1959) No. 1 (8), 156-164 (Mathnet URL: \url{http://mi.mathnet.ru/eng/ivm2415}). [in Russian: On a complex representation of the tensors of Lorentz space].
\bibitem{Penrose1}
R. Penrose, W. Rindler: Spinors and Space-Time. Vol. 1: Two-Spinor Calculus and Relativistic Fields. Vol. 2: Spinor and Twistor Methods in Space-Time Geometry. Cambridge Monogr. Math. Phys.. Cambridge University Press, Cambridge, Vol. 1: 1984, Vol. 2: 1986. Russian translation: Р. Пенроуз [R. Penrouz], В. Риндлер [V. Rindler]: Спиноры и пространство-время [Spinory i prostranstvo-vremya]. Мир [Mir], Москва [Moskva], т. 1 [Tom 1]: 1987, т. 2 [Tom 2]: 1988. Russian translation by Е.М. Серебрянный [E.M. Serebryanny\u\i] and З.А. Штейнгард [Z.A. Shеу\u\i ngard] under edition Д.М. Гальцов [D.M. Gal'tsov].
\bibitem{Petrov1}
А.З. Петров [A.Z. Petrov]: Классификация пространств, определяющих поля тяготения [Klassifikatsiya prostranstv, opredelyayushchikh polya tyagoteniya]. Уч. Зап. Казан. Гос. Унив. [Uch. Zap. Kazan. Gos. Univ.] 114(1954) No. 8, 55-69 (Mathnet URL: \url{http://mi.mathnet.ru/uzku344}). English translation: A.Z. Petrov: The classification of spaces defined by gravitational fields. Gen. Rel. Grav. 22(2000)1665-1685 (\href{http://dx.doi.org/10.1023/A:1001910908054}{DOI: 10.1023/A:1001910908054}).
\bibitem{Rosenfeld1}
Б.А. Розенфельд [B.A. Rozenfel'd]: Проективная геометрия как метрическая геометрия [Proektivnaya geometriya kak metricheskaya geometriya]. Труды семинара по векторному и тензорному анализу с их приложениями к геометрии, механике и физике [Trudy seminara po vektornomu i tenzornomu analizu s ikh prilozheniyami k geometrii, mekhanike, fizike], Москва [Moskva], 8(1950)328-354, [in Russian: Projective geometry as the metric geometry. Proceedings of the Seminar on Vector and tensor analysis with their applications.]
\bibitem{Rosenfeld2}
Б.А. Розенфельд [B.A. Rozenfel'd]: Неевклидовы геометрии [Neevklidovy geometrii]. ГИТТО [GITTO], Москва [Moskva], 1955. [in Russian: Non-Euclidean Geometries]. The Cartan triality principle is given on the page 534.
\bibitem{Rosenfeld3}
Б.А. Розенфельд [B.A. Rozenfel'd]: Многомерные пространства [Mnogomernye prostranstva]. Наука [Nauka], Москва [Moskva], 1966. [in Russian: Multidimensional Spaces].
\bibitem{Scharnhorst1}
K. Scharnhorst  and J.-W. van Holten: Nonlinear Bogolyubov-Valatin transformations: 2 modes. Annals of Physics (New York), 326(2011)2868-2933 [\href{http://arxiv.org/abs/1002.2737}{arXiv:1002.2737v3}, NIKHEF preprint NIKHEF/2010-005] (\href{http://dx.doi.org/10.1016/j.aop.2011.05.001}{DOI: 10.1016/j.aop.2011.05.001}
\bibitem{Sintcov1}
Д.М.Синцов [D.M. Sintsov]: Теория коннексов в пространстве в связи с теорией дифференциальных уравнений в частных производных первого порядка. [Teoriya konneksov v prostranstve v svyazi s teorie\u\i\ differentsial'nykh uravneni\u\i\ v chastnykh proizvodnykh pervogo poryadka]. Типо-литография Императорского Университета [Tipo-litografiya Imperatorskogo Universiteta], Казань [Kazan'], 1894. [in Russian: Theory of connexes in space in relation to the theory of first order partial differential equations].
\bibitem{Stepanovskii1}
Ю.П. Степановский [Yu. P. Stepanpovski\u\i]: Спиноры 6-мерного пространства и их применение к описанию поляризованных частиц со спином 1/2 [Spinory 6-mernogo prostranstva i ikh primenenie k opisaniyu polyarizovannykh chastits so spinom 1/2]. Проблемы Ядерной Физики и Космических Лучей [Problemy Yaderno\u\i\ Fiziki i Kosmicheskikh Luche\u\i], Харьков [Khar'kov], 4(1976) 9-21, [in Russian: Spinors of a sixdimensional space and their application to the description of polarized particles with spin 1/2].
\bibitem{Stepanovskii2}
Ю.П. Степановський [Yu. P. Stepanpovski\u\i]. Алгебра матриць Дiрака у шестiмiрному виглядi [Algebra matrits Diraka u shestimirnomu vyglyadi]. Украiнський Фiзичний Журнал [Ukrains'ky\u\i\ Fizychny\u\i\ Zhurnal]. Kиiв [Kyiv], т. XI, 8(1968) 813-824, in Ukraine: Yu. P. Stepanovskii: Algebra of Dirac matrices in six-dimensional form. Ukrainian Journal of Physics, Kiev, v.XI, 8(1968) 813-824.
\bibitem{Tarski1}
Alfred Tarski. Introduction to Logic and to the Methodology of Deductive Sciences. New York, Oxford, OXFORD UNIVERSITY PRESS, 1994. Russian translation: Альфред Тарский [Al'phred Tarski\u\i]. Введение в логику и методологию дедуктивных наук [Vvedenie v logiku i mrtodologiyu deduktivnykh nauk]. ГИИЛ [GIIL], Москва [Moskva], 1948. Russian translation by О.Н. Дынник [O.N. Dynnik].
\bibitem{Rozenfeld4}
Хуа Ло-гэн [Khua Lo-g\`en], Б.А. Розенфельд [B.A. Rozenfel'd]: Геометрия прямоугольных матриц и ее применение к вещественной проективной и неевклидовой геометрии.
[Geometriya pryamougol'nykh matrits i ee primenenie k veshchestvenno\u\i\ proektivno\u\i\ i neevklidovo\u\i\ geometrii]. Изв. Высш. Учебн. Завед., Матем. [Izv. Vyssh. Uchebn. Zaved., Matem.] (1957) No. 1, 233-247 (Mathnet URL: \url{http://mi.mathnet.ru/ivm3038}). English translation: Hua Loo-geng (Hua Loo-keng), B.A. Rozenfel'd: The geometry of rectangular matrices and its application to real-projective and non-euclidean geometry. Chin. Math. 8(1966)726-737.
\bibitem{Zee1}
A. Zee: Quantum field theory in a nutshell. Princeton university press, 2003. Russian translation: Зи Энтони [Zi \`Entoni]. Квантовая теория в двух словах [Kvantovaya teoriya v dvukh slovakh]. РХД [RKHD], Москва [Moskva], Ижевск [Izhevsk], 2009. Russian translation by В.Г. Войткевич [V.G. Vo\u\i tkevich] and Ю.В. Колесниченко [Yu.V. Kolesnichenko] under edition И.В. Полюбина [I.V. Polyubina].

}
\end{document}